# Distributed Intrusion Detection System using Semantic-based Rules for SCADA in Smart Grid

Sathya Narayana Mohan, Gelli Ravikumar, *Member, IEEE*, and Manimaran Govindarasu, *Fellow, IEEE*

*Abstract*—Cyber-physical system (CPS) security for the smart grid enables secure communication for the SCADA and wide-area measurement system data. Power utilities world-wide use various SCADA protocols, namely DNP3, Modbus, and IEC 61850, for the data exchanges across substation field devices, remote terminal units (RTUs), and control center applications. Adversaries may exploit compromised SCADA protocols for the reconnaissance, data exfiltration, vulnerability assessment, and injection of stealthy cyberattacks to affect power system operation. In this paper, we propose an efficient algorithm to generate robust rule sets. We integrate the rule sets into an intrusion detection system (IDS), which continuously monitors the DNP3 data traffic at a substation network and detects intrusions and anomalies in real-time. To enable CPS-aware wide-area situational awareness, we integrated the methodology into an open-source distributed-IDS (D-IDS) framework. The D-IDS facilitates central monitoring of the detected anomalies from the geographically distributed substations and to the control center. The proposed algorithm provides an optimal solution to detect network intrusions and abnormal behavior. Different types of IDS rules based on packet payload, packet flow, and time threshold are generated. Further, IDS testing and evaluation is performed with a set of rules in different sequences. The detection time is measured for different IDS rules, and the results are plotted. All the experiments are conducted at Power Cyber Lab, Iowa State University, for multiple power grid models. After successful testing and evaluation, knowledge and implementation are transferred to field deployment.

## I. INTRODUCTION

THE modern smart grid, profoundly relies on Ethernet-based communication protocol like DNP3. The central aspect of smart grid is to create a decentralized power distribution where the consumer involves towards power contribution locally [1]. Network data exchanged through SCADA communication has high importance to availability, integrity, and confidentiality. In this order, availability is extremely critical as all the processes operates in real-time. In traditional security, confidentiality and integrity of the message are given more priority. However in critical communications, availability of data packets are given primary importance [2], then follows integrity. Any slight compromise towards integrity of data may lead to disastrous events. Confidentiality has lowest priority in comparison of others. Since encryption creates an additional overhead time, which hinders the availability of critical communication. The data communication in most of the SCADA environment are not encrypted.

This work focuses on providing distributed IDS/IPS to decrease an attack surface of a system by using robust set of IDS rules. It describes distributed intrusion and prevention system architecture and explains in detail about network nodes and operation. Cyber attacks, which are developing more stealthy [3] [4] demands rule-based, analysis-based, and behavior-based detection to prevent real-time attackers. There is also demand in monitoring the network activity, like analysis of network behavior of the system which has anomalous patterns. In upcoming section, types of traffic pattern are studied. The IDS rule generation algorithm is discussed considering different types of properties and Cyber incidents. At the same time, different types of SCADA traffic are observed and recorded. Then, several types of IDS rules are generated based on the given scope of cyber attacks. An overview of Cyber attack classification is presented and an evaluation of the detection system is performed to show how the order of IDS rules can decrease detection time.

## II. DISTRIBUTED IDS FOR LARGE-SCALE SMART GRID

SCADA Smart grid relies extensively on the poll and response communication of critical controls and data objects. The communication that happens between different networks in the grid uses Distributed network protocol (DNP3). Gaining an understanding of network activities are required in real-time to protect against cyber attacks. This section shows the elements of a distributed IDS with Intrusion prevention system built using In-line architecture [5]. Distributed IDS architecture, includes a master node and many sensors. All clients are represented as forward nodes and are connected to master node. Each client has a sensor installed that operates in promiscuous mode to collect network information and forwards to master node. In Standalone operation, master node has its own database server. Whereas, distributed deployment has two options with or without storage node. In this type, master node maintains a storage node to forward network data information for future queries. Further, distributed IDS is loaded with IDS rules that are designed based on traffic pattern, network packets, packet flow, packet content and packet threshold time. This potential feature can be tailored for different types of network traffic environment. It gives enormous flexibility for a security expert to analyze and monitor the network data from a single standpoint [6]. Also, an operator can leverage the flexibility of writing different types of IDS rules that suits accordingly.

### A. Distributed Deployment

The Distributed deployment comprises of a master node, many forward and storage nodes. This architecture is widely deployed in production as it provides endpoint monitoring. It also handles heavy network traffic and log management using dedicated storage nodes. It is highly recommended to use distributed deployment for production network, and the use of storage node provides more extended options to learn about data analytic and elastic search [6].

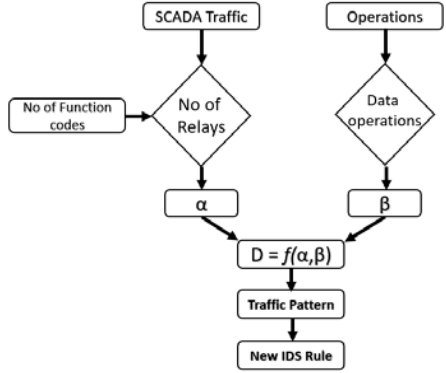

Fig. 1. Design flow to detect traffic pattern

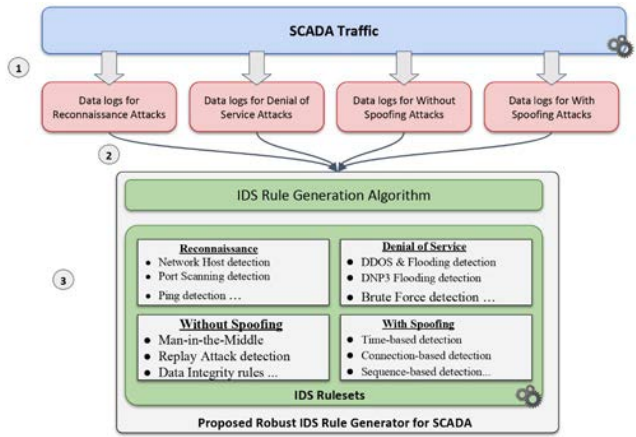

Fig. 2. Rule Generation based on Cyber incidents

### B. In-line IPS Deployment

The IDS uses signatures to match network packets for finding an alert [7]. When an IDS is placed at the gateway of a network or in-line to device, it can now operate as a prevention system. The intrusion prevention system is able to detect and defend against various types of cyber attacks like Denial of service. This type of implementation secures network as it provides functionality of detection and prevention of malicious activities.

## III. METHODOLOGY

### A. IDS Rule Generation Algorithm

In this section, we discuss IDS rule generation. Here, we define different sets of IDS rule based on various traffic patterns. The primary step in IDS rule generation is to study network traffic of SCADA environment. In this case, we consider DNP3 traffic at Power Cyber Lab, Iowa state university [8] to analyze different commands that are exchanged between control center and substations. The communication of DNP3 commands follows a particular pattern that depends on utility conditions. Based on the features listed in Figure 3 traffic patterns are classified as in Figure 4.

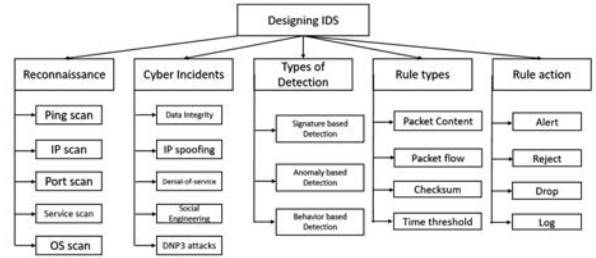

Fig. 3. Properties to Design Robust IDS

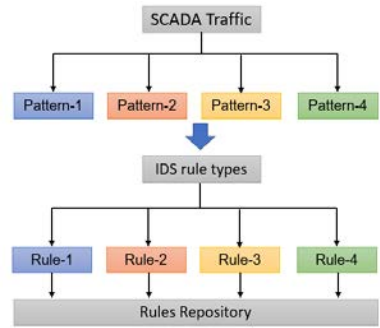

Fig. 4. Rule pattern

The Figure 1 shows design flow to identify a traffic pattern, which involves two functions $\alpha$, number of relays with the number of functions per relay and $\beta$, data operations with respect to time. The Figure 2 shows different traffic patterns to consider for creating robust IDS rules on various cyber incidents. In this rule design, we consider data logs for basic classification of cyber attacks and generate rule algorithm to detect the incidents. After considering, different traffic patterns are classified based on packet content, packet flow, function code, and time period. From the Figure 4 in pattern-1, we consider critical ports and services used with respect to packet payload. In pattern-2, we consider packet flow concerning server and substation. In DNP3, a request message is sent from Master to substation and a response message is sent from substation to Master. In pattern-3, we consider a critical commands with respect to time period. Some critical commands like select and operate are time sensitive. In pattern-4, the time period is measured for different network packets. The traffic pattern can be based on protocol, services, critical commands, packet payload, content, time threshold. From this consideration, a list of traffic patterns are defined. After having the traffic pattern, the IDS rules are tailored to detect a particular incident. In the Fig. 4, it shows that when a new pattern is detected, a new IDS rule based on different properties is generated. After a new IDS rule is generated, the rule is sent to repository and it is updated with regular network operation. This brings a robust adaptive approach to detect different types of cyber incidents in a network. Similarly, each Substation has many Relays, and each relay may control many endpoints. The traffic pattern for each end device is different. Now, each relay has different endpoints and are controlled by different function codes. Therefore, the network traffic is different for each end-device of different relays and substations. Hence, traffic patterns are separated based on station IDs, relay IDs, and device addresses. Fig. 4 shows IDS rule design flow. Fig. 3 shows the functions to consider for creating robust IDS rules for various endpoints.

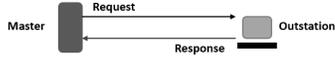

Fig. 5. DNP3 communication between Master and Substation

| Code | Function | Code | Function |
|---|---|---|---|
| 00 | Confirm | 10 | Initialize application |
| 01 | Read | 11 | Start application |
| 02 | Write | 12 | Stop application |
| 03 | Select | 13 | Save configuration |
| 04 | Operate | 14 | Enable unsolicited |
| 05 | Dir operate | 15 | Disable unsolicited |
| 06 | Dir operate-No resp | 16 | Assign class |
| 07 | Freeze | 17 | Delay measurement |
| 08 | Freeze-No resp | 18 | Record current time |
| 09 | Freeze clear | 19 | Open file |
| A | Freeze clear-No resp | 1A | Close file |
| B | Freeze at time | 1B | Delete file |
| C | Freeze at time-No resp | 1C | Get file information |
| D | Cold restart | 1D | Authenticate file |
| E | Warm restart | 1E | Abort file |
| F | Initialize data | | |

Fig. 6. DNP3 Request messages

## IV. DNP3 PROTOCOL

DNP3, an open standard communication protocol that supports communication to multiple devices at different outstation levels. DNP3 communication also supports multiple data types such as critical-control data, time-synchronized data, time-stamped data, data with priorities, data with responses, data with acknowledgments. It handles two-way request-response communication. The communication provides unicast, multicast, anycast, and broadcast, which gives protocol to handle distributed architecture seamlessly.

In DNP3 communication, as shown in the Fig. 5, the master and outstation exchanges two types of messages namely; request and response. The master sends a request message to its respective outstation to enquire about status of outstation. Similarly, outstation replies a response message with its status information to the master. The Figure 6 shows the list of request message from master to outstation. It provides various function codes to perform different operations using DNP3 request.

## V. CYBER ATTACK CLASSIFICATION

The Figure 7 shows scope of cyber attacks considered in this study. The incidents are categorized based on four types, as shown in the figure. Reconnaissance includes all kinds of system scanning like IP addresses, ports, services, operating systems. Denial-of-service, one of the frequent incidents that occur in any network. It includes all types of packet flooding such as Ping-to-death, IP flooding, DNP3 flooding, Distributed

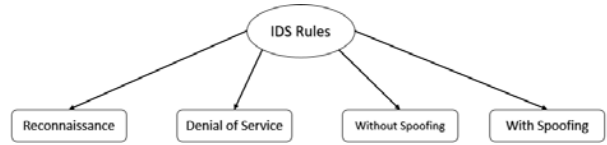

Fig. 7. Classification of Cyber attacks

Denial of service (DDOS), brute force attack [9]. The Classification is then divided into with and without spoofing. In spoofing attacks, an attacker uses a spoofed IP header and appends DNP3 payload to bypass security measures and hit the target. However, based on traffic analysis, it is able to detect spoofing based attacks. In the following section, there are IDS rules which can identify these spoofing attacks are discussed.

### A. Common attacks in DNP3 protocol stack

In this work, critical commands are used to execute an attack and observe the instability of the system. These critical commands directly operate the remote system when a scripted packet is delivered. There are many types of data integrity attacks related to TCP/IP layer [4]. In this work, we study the following DNP3 data integrity attacks. All attacks mentioned here have experimented with real-time testing using python scripts, and a suitable IDS rule is designed to prevent it. The Cyber attack and defense work is conducted, tested, and the rule detection performance is analyzed in Power cyber lab, Iowa state university.

*1) DNP3 select-operate attack:* In DNP3, Select and Operate are two Functions. During system operation in SCADA, Select command is always initiated first and then Operate. The attacker sniffs payload of both commands and delivers it to the target. This attack can lead to disastrous situations like a blackout, power outage.

*2) DNP3 direct-operate attack:* As compared to previous attack, it is more severe. Here the function does not involve with the Select, and it directly operates the system. Now, it becomes simple for an attacker to modify one payload instead of two. It can affect the power system and leads to a blackout. An IDS rule with content and time threshold is given to check whether traffic pattern is normal.

*3) DNP3 broadcast request attack:* In this type of attack, the attacker modifies the destination station address in the DNP3 link layer to broadcast address. This attack broadcasts the critical function code to all the station addresses in the network and affects the system. This attack can lead to multiple impacts at different substations.

*4) DNP3 disable unsolicited responses:* In Unsolicited response, an outstation sends a response without any request from the server if there is any unexpected change in the system status. In this type of attack, the attacker sends disable unsolicited response to the outstation. It intercepts communication between server and substation, which makes the server unaware of any unexpected operation. This attack can lead to a situation where a system can be separated from real-time communication. A python script is made to test and an IDS rule is designed to alert it.

| Rule type | Sample rule |
|---|---|
| Packet payload | alert tcp !$SRC any -> $DST any (content:"|04|"; offset:12; depth:1; msg:"DNP3 operate from Unknow source"; sid:3;) |
| Packet flow | alert tcp !$SRC any -> $DST any (flow: not_established; msg:"Unknown flow"; sid:5;) |
| Time-threshold | alert tcp !$SRC any -> $DST any (content:"|05 64|"; threshold: type both, track by_src, count 5, seconds 10; sid:9;) |
| Incorrect Checksum | alert tcp !$SRC any -> $DST any (msg:"DNP3-Bad-CRC"; sid:1; gid:145; metadata: rule-type preproc;) |
| Invalid TCP sequence number | alert tcp !$SRC any -> $DST any (msg:"DNP3-Invalid sequence no"; sid:3; gid:145; metadata: rule-type preproc;) |

Fig. 8. IDS rule sample

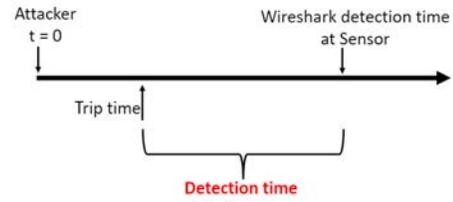

Fig. 9. IDS rule detection time

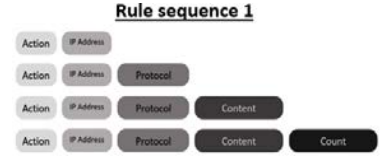

Fig. 10. IDS Rule Sequence 1

*5) DNP3 stop application attack:* The function code related to Stop application is 12; an attacker can use this information to exploit and perform unauthorized action. In Stop application attack, an attacker tries to shut down or stop a functioning application. The impact of this can be instability in the power system or any control systems. An IDS rule with content field is used to counter it.

*6) DNP3 Cold restart request:* Cold restart is used to clear the communication sequence and restart an end device. This is generally used to check self-test of the device. The function code related to Cold restart is D (in hexadecimal); an attacker attempts this to perform an unauthorized operation. An IDS rule with a time threshold is employed to check the number of packets and prevent it.

## VI. DIFFERENT TYPES OF IDS RULES

After classifying different patterns of traffic, based on different features as discussed in the previous section. We now discuss, types of IDS rules to detect above types of attacks. Using network traffic pattern and rule option fields an IDS rule set is designed as shown in the Figure 8.

### A. Rules based on Network Packet Payload

This rule enables the detection if an attacker injects any corrupted packet having critical function to execute a system. In this rule, only communication between DNP3 server and Home network is allowed. The IP address of DNP3 server and Home network are predefined at master and sensor system. The rule options field has both offset and depth of the packet to identify which function code used. This rule detects data integrity attacks that happens at the DNP3 layer.

### B. Rules based on Network Packet flow

It provides an option of checking whether packet flow is established or not established. Before any TCP session is established, there exists a three-way handshake between client and server. This rule checks if the packet flow between the two sides are established or not established. Depending on the source and destination IP address, port number, and the packet payload rule is made. All TCP connection starts with SYN, SYN-ACK, and ACK; this rule alerts when there is any exchange of three messages. It also detects all types of scanning of IP address, port scanning, OS and service scanning.

### C. Rules based on Time-threshold

This rule is written based on time threshold of DNP3 packet. It checks the network traffic, whether it is exceeding the number of attempts of a particular packet. If there is any attempt of exceeding a DNP3 packet, an alert is triggered. This rule detects all types of flooding attacks that happen in the SCADA environment. Flooding attacks like TCP flooding, SYN flooding, MAC flooding, reverse TCP flooding, scanning, DNP3 flooding, Denial-of-service, Distributed denial-of-service are detected.

### D. Rules based on Incorrect Checksum

The checksum of a network packet shows whether transmitted data has reached without any errors. This rule detects whether the checksum is matching at the receiving end. An alert is generated if an incorrect checksum is found. The rule option field has gid (generator id); these are predefined identifiers in the rule detection engine to check the checksum of the network packet.

## VII. EVALUATION AND RESULTS

Figure 9 shows how detection time is calculated. The trip time from the attacker machine to the RTU is measured, the difference between Wireshark time at sensor and trip time gives detection time. The IDS detection time is calculated for all the rules in this fashion. Figure 10 shows first rule order sequence. In this sequence, rules are arranged uniformly based on the action, protocol, content, and count. The detection time for each of the rule is noted and plotted in the Figure 11.

Similarly, Figure 12 shows second rule order sequence, here rules are arranged in reverse order. The order is based on the count, content, protocol, and action to check detection time and performance. The detection time for each of rule in Sequence-2 is recorded and plotted in the Figure 13.

The Figure 14 shows a comparison between two rule sequences. The detection time for Ping, packet flow and content rule in sequence-1 is lesser than sequence-2. Whereas, detection time for time threshold rule in sequence-2 is lesser than sequence-1. This infers that processing time is less for rules placed on top order compared to others. Thus, critical IDS rules are placed on top order of rule repository to get minimum detection time. The detection time also depends

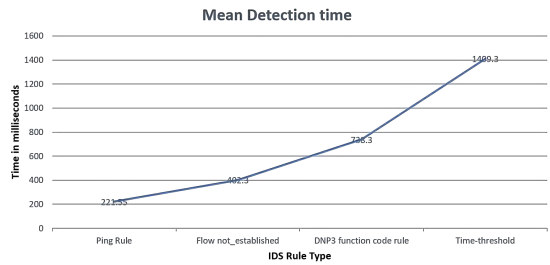

Fig. 11. Detection time for Sequence 1

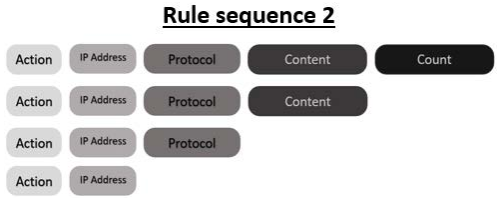

Fig. 12. IDS Rule Sequence 2

on the rule length, if a rule contains many options field, the processing time increases.

## VIII. FIELD DEPLOYMENT

In this section, we discuss on how this work was successfully transferred to a local power utility in Iowa. For security reasons, the utility name is not disclosed. The deployment consisted of two phases. In the first-phase, master and client-1 are deployed. In the second-phase, client-2 is deployed. This work showed an advantage of having distributed IDS and the effectiveness of customized IDS rules. After conducting repeated experiments for different IDS rules based on various scenarios, both phases were completed. The power utility has a primary control center, where all the data operations are conducted. The control center has a network of substation which are connected in ring fiber link. The robust rule set developed based on this work is transferred in Phase I. The rule set is updated at master; then pushed to client-1 at substation-1 virtually.

## IX. CONCLUSION

Securing data communications is one of the essential aspects to consider in critical infrastructures. While no network is immune to attacks, but a stable and an effective network security is necessary to protect industrial communications. There are several layers of network-level protection to prevent cyber attacks like Man-in-the-middle, Data integrity attacks, DOS and Social engineering attacks. This paper describes a typical approach in IDS/IPS by making it distributed to monitor the network activity of different networks. Several attack signatures were formulated and executed to study the behaviour of system processes. To prevent this, suitable IDS rules were designed and their operation is tested. The design of the rule generation algorithm is explained based on different SCADA traffic patterns. In these traffic pattern, various IDS rules are designed and the implementation is deployed in distributed architecture. The cyber attack classification is prepared within the scope of lab, and testing is performed for designed IDS rules. Then, an evaluation of IDS rules is completed to check minimum detection time. Finally, experimental results are collected for two different rule order sequences and found that the processing time is merely less for the rules placed on top order of repository. From this observation, it is perceived that IDS rules which are having high impact are kept on top order. Further, field deployment was completed successfully and the possible extension of this work would be introducing data analytic, with new machine learning techniques to develop advanced detection.

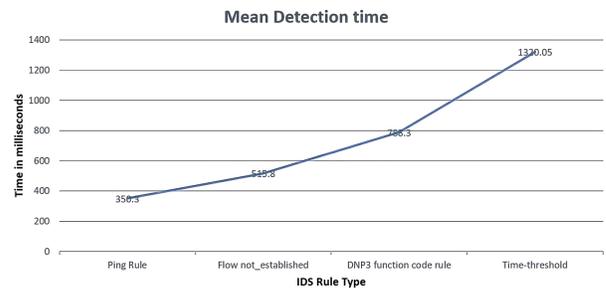

Fig. 13. Detection time for Sequence 2

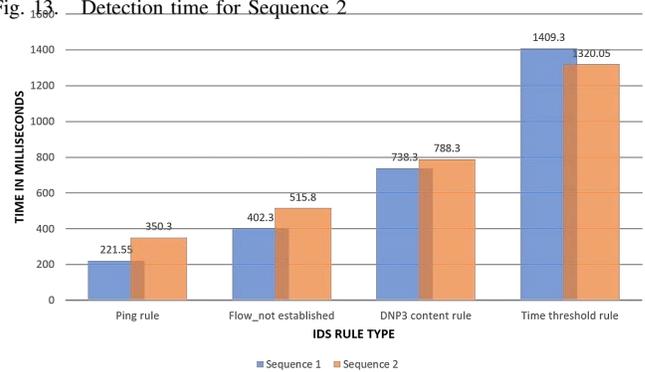

Fig. 14. Sequence 1 vs Sequence 2

## X. ACKNOWLEDGEMENT

This research is funded in part by US NSF Grant # CNS 1446831, and US DOE Grant # DE-OE0000830.